\documentclass[pdftex,twocolumn,epjc3]{svjour3}          

\RequirePackage[T1]{fontenc}

\smartqed  
\usepackage{array} 
\RequirePackage{amsmath,amssymb}
\RequirePackage{bm}
\RequirePackage{booktabs}
\RequirePackage{graphicx}
\RequirePackage{mathptmx}      
\RequirePackage{flushend}
\RequirePackage{microtype}     
\RequirePackage[numbers,sort&compress]{natbib}
\RequirePackage[colorlinks,citecolor=blue,urlcolor=blue,linkcolor=blue]{hyperref}
\RequirePackage{placeins}
\RequirePackage{stfloats}
\usepackage{comment}
\makeatletter
\renewcommand\makeheadbox{}   
\patchcmd{\@maketitle}{\hrule\@height0.35mm}{\hrule\@height\z@}{}{}  
\makeatother
\begin{document}

\title{Ground-State Heavy-Baryon Mass Spectrum in the Relativistic Independent Quark Model}

\author{Divyajyoti Pandey\thanksref{e1,addr1,addr2}
        \and
        Kalpalata Dash\thanksref{addr3}
        \and
        Sanjay Kumar Swain\thanksref{addr1,addr2}
}

\thankstext{e1}{e-mail: divyajyoti.pandey@niser.ac.in}

\institute{National Institute of Science Education and Research, Bhubaneswar,
Khordha, Odisha - 752050, India.\label{addr1}
          \and
          Homi Bhabha National Institute, Training School Complex, Anushakti Nagar,
Mumbai - 400094, India.\label{addr2}
\and
Prananath College (Autonomous), Khordha, Odisha - 752057, India.\label{addr3}
}

\date{}
\maketitle
\begin{abstract}

We extend our work to compute the ground-state baryon mass spectrum containing charm and bottom quarks in the relativistic independent quark (RIQ) model with an average flavor--independent confining potential in equally mixed scalar--vector harmonic form. Each quark moves as a Dirac particle inside the potential. The resulting core mass is then corrected perturbatively, accounting for the pion self-energy, pion exchange, and pairwise color--electric and color--magnetic one-gluon exchange. Six input parameters common to every baryon: the potential parameters (confinement strength, the potential offset) and the $u(=d)$, $s$, $c$, and $b$ constituent quark masses are optimized using a few experimentally measured baryon states. The model then uses these optimized parameters to predict the mass spectra of doubly heavy, mixed-$bc$, and triply heavy baryons. We compare our predictions with lattice QCD, relativistic quark model results, and available experimental measurements. Our results are consistent with these model predictions and experimental results. 
\end{abstract}

\section{Introduction}
\label{sec:intro}
In particle physics, the classification of baryons and mesons in terms of quarks~\cite{Gell-Mann:1964ewy, Zweig:1964jf} and the role of color as the charge of the confining gauge
field~\cite{Fritzsch:1973pi} laid the foundation for understanding the hadron spectrum. The available data from experiments on charmed and bottom baryons~\cite{
ParticleDataGroup:2024cfk} have made it possible to study light and heavy quarks within a single relativistic framework.
Several non-relativistic and relativistic quark potential models have attempted to study the meson spectrum, both light and heavy-light~\cite{Vijande:2004he, Shlyapnikov:2000iw,
Bhavyashri:2005rb,Li:2019tbn}, and heavy quarkonia system~\cite{Ebert:2009ua,Godfrey:2004ya,Segovia:2013wma,PhysRevD.51.6348,Solovev:1997htw,Fischer:2014cfa,Chen:2020ecu}, along with the increasingly precise lattice-QCD predictions of the meson spectrum~\cite{Thomas:2017rkj}. For baryons, Faddeev and
quark-cluster calculations that solve the genuine three-body problem have been used for the light and strange sectors~\cite{Julia-Diaz:2001pxb, Garcilazo:2001ck, Julia-Diaz:2001apa,
PhysRevC.59.428}. However, the present calculation follows a different, computationally simpler method: the independent-quark approximation, with translational invariance restored afterwards by the projection technique mentioned later in the paper.

In the original RIQ model of Barik and Collaborators, each confined quark obeys a Dirac equation in an equally mixed scalar--vector harmonic potential, and the resulting independent-quark core is supplemented perturbatively by pion and gluon exchange and by a correction for the center-of-mass motion~\cite{Barik:1986mq, Barik:1987zb}. The model has been used to compute baryon properties and electroweak form factors~\cite{Barik:1985rm, Barik:1985in}, the weak leptonic decays of light and heavy pseudoscalar mesons~\cite{Barik:1993aw}, and more recently the study of 
a range of leptonic, exclusive semileptonic and nonleptonic $B_{(d,c,s)}$--decays~\cite{Dash:2023ohf, Nayak:2021djn, Nayak:2022qaq, Patnaik:2023efe, Nayak:2024esq, Patnaik:2023ins, Dash:2024crn, Dash:2023hjr}. A similar study of the full meson and baryon ground-state spectra has also been studied using the chiral quark model~\cite{He:2023gqh}. We adopted this general RIQ framework but extended it to baryons containing charm and bottom quarks, including the doubly heavy, mixed-$bc$, and triply heavy configurations. 

We treat the baryon throughout as a three-quark state: we project the full localized product of three ground-state Dirac orbitals onto zero total momentum using the numerical Peierls--Yoccoz (PY) projection method~\cite{Peierls:1957zz}, then add pion self-energy, pion exchange, and pairwise color--electric and color--magnetic corrections. A single set of six input parameters common to all baryons is optimized as in the original construction. The method used in this paper closely follows the original Barik--Dash RIQ construction and remains valid across the much wider light-to-triply-heavy mass range discussed in the paper. The steps are: (i) solve the Dirac equation for each quark flavor; (ii) build the color--singlet three-quark state with the appropriate spin-pair coefficients; (iii) project onto zero momentum to obtain the baryon core mass; (iv) calculate the pionic and gluonic corrections; (v) sum the contributions; (vi) find the optimized model parameters $(a,V_0,m_u = m_d,m_s,m_c,m_b)$ to predict the baryon masses.

In this paper, Sec. 2 provides detailed information on the RIQ model framework and the general kinematics, including the single-quark Dirac equation, the center-of-mass correction and residual interactions used for the calculation. Sec. 3 presents the results and discussion on the baryon mass predictions and comparison with experimental values, along with predictions from other theoretical models. Finally, Sec. 4 provides a summary and conclusions.

\section{Relativistic Independent Quark (RIQ) Model}
\label{sec:core}

\subsection{Single-quark Dirac equation}
\label{sec:singlequark}
In the RIQ model, each constituent quark of mass $m_q$ moves in the mean-field Hamiltonian of the form:
\begin{equation}
H_q = \bm{\alpha}\cdot\mathbf{p} + \beta m_q + U(r), \qquad
U(r) = \tfrac{1}{2}(1+\beta)\left(ar^2+V_0\right)
\label{eq:H1}
\end{equation}
where $\mathbf p$ is the quark's momentum operator, $\bm\alpha=(\alpha_1,\alpha_2,\alpha_3)$ and $\beta$ are the standard Dirac matrices, $a$ is the confinement strength and $V_0$ is a constant potential offset -- both parameters together fix the shape of the harmonic well $U(r)$. This Hamiltonian, with $a > 0$, is believed to provide zeroth-order constituent
quark dynamics inside the hadron bound state, and $\bm\alpha,\beta$ obey the algebra $\{\alpha_i,\alpha_j\}=2\delta_{ij}$,
$\{\alpha_i,\beta\}=0$, $\alpha_i^2=\beta^2=1$. The wave function is a four-component spinor that is known from the standard Dirac equation built from the Pauli matrices $\sigma_i$. The potential is an equal admixture of a Lorentz-scalar confinement $S(r)=\tfrac12 (ar^2+V_0)$, added to the mass term through $\beta$, and a time-like Lorentz-vector confinement $V(r)=\tfrac12 (ar^2+V_0)$ of the same strength. With the substitution of
\begin{equation}
E_q' = E_q-\tfrac{V_0}{2},\quad m_q'=m_q+\tfrac{V_0}{2},\quad \lambda_q=E_q'+m_q'
\label{eq:H2}
\end{equation}
the eigenvalue equation $H_q\psi=E_q\psi$ gives the solutions for the quark wave function
\begin{equation}
\psi_{n\kappa m}(\mathbf r)
=
\begin{pmatrix}
\displaystyle i\,\frac{g_{n\kappa}(r)}{r}\,
\Omega_{\kappa m}(\hat r)
\\[10pt]
\displaystyle -\,\frac{f_{n\kappa}(r)}{r}\,
\Omega_{-\kappa m}(\hat r)
\end{pmatrix}
\label{eq:spinoransatz}
\end{equation}
Here $g_{n\kappa}(r)$ and $f_{n\kappa}(r)$ are, respectively, the large (upper) and small (lower) radial wave functions, $n$ is the radial quantum number, $\Omega_{\kappa m}$ are the usual spherical spinors, $m$ the magnetic quantum number, and $\kappa$ is the relativistic (Dirac) quantum number that fixes both the total angular momentum $j=|\kappa|-\tfrac12$ and the orbital angular momenta of the upper and lower components. $\kappa=-1$ labels the ground-state $1s_{1/2}$ level ($j=\tfrac12$, upper orbital angular momentum $l=0$, lower orbital angular momentum $l'=1$), the only state needed here since only ground-state baryons are considered in the paper. With $S(r)=V(r)$, Eq.~\eqref{eq:spinoransatz} reduces to the coupled radial Dirac equations
\begin{align}
\frac{dg}{dr}+\frac{\kappa}{r}g &= \left(m_q+ar^2+E_q'-\tfrac{V_0}{2}\right)f \\
\frac{df}{dr}-\frac{\kappa}{r}f &= \left(m_q+ar^2-E_q'+\tfrac{V_0}{2}\right)g
\label{eq:radialpair}
\end{align}
Since the confining potential is quadratic in nature, the ground state has the Gaussian solution of the form
\begin{equation}
g(r) = N\,\frac{r}{r_{0q}}\,e^{-r^2/2r_{0q}^2}, \qquad
f(r) = N\eta_q\left(\frac{r}{r_{0q}}\right)^{2} e^{-r^2/2r_{0q}^2}
\label{eq:gaussianansatz}
\end{equation}
with normalization $N$, oscillator length $r_{0q}$, and lower component to upper component amplitude ratio
$\eta_q$. After substituting the above solutions into Eq.~\eqref{eq:radialpair} fixes $r_{0q}=(a\lambda_q)^{-1/4}$
and $\eta_q=1/(r_{0q}\lambda_q)$, while requiring the full radial equation to be satisfied for
the ground-state combination $s\equiv4n+2l-1=3$ gives the algebraic condition
\begin{equation}
\left(E_q'-m_q'\right)^2\left(E_q'+m_q'\right)=9a
\label{eq:cubic}
\end{equation}
a cubic equation for $E_q'$ that can be  solved by Newton--Raphson method for each quark flavor $q=u,d,s,c,b$, with $m_d\equiv m_u$. The lower component of the
solution is appreciable for the light quarks as they are genuinely relativistic, but is progressively suppressed for $c$ and $b$, as expected for general heavy-quark reasons.

Now, the baryon wave function factorizes as
\begin{equation}
\Psi_B = \Phi_{\rm space}\,\chi_{\rm color}\,\chi_{\rm spin}\,\chi_{\rm flavor}\,
\end{equation}
where $\Phi_{\rm space}=\prod_i\psi_i$ is the product of the three single-quark orbitals mentioned in  Eq.~\eqref{eq:spinoransatz}, and $\chi_{\rm spin}$, $\chi_{\rm flavor}$ are the total spin and flavor wave functions. The color wave function is taken to be the normalized antisymmetric color-singlet state, $\chi_{\rm color}=\epsilon_{abc}|abc\rangle/\sqrt6$, where $\epsilon_{abc}$ is the Levi-Civita symbol in color space.
Since $\chi_{\rm color}$ is
antisymmetric and $\Phi_{\rm space}$ is symmetric in the ground state, the spin--flavor part must be symmetric under interchange of identical quarks. The pairwise spin expectation value is
\begin{equation}
\langle\bm\sigma_i\!\cdot\!\bm\sigma_j\rangle = 2S_{ij}(S_{ij}+1)-3
= \begin{cases} -3, & S_{ij}=0\\ +1, & S_{ij}=1\end{cases}
\label{eq:sigmadot}
\end{equation}
where $S_{ij}$ is the total spin ($0$ or $1$) to which the two spin-$\tfrac12$ quarks $i,j$ are coupled,
and an analogous isospin factor $\langle\bm\tau_i\!\cdot\!\bm\tau_j\rangle=2I_{ij}(I_{ij}+1)-3$,
with $I_{ij}$ the total isospin of the quark pair, similarly built from the isospin Pauli matrices $\bm\tau_i$,
that appears in the pion-exchange term of Eq.~\eqref{eq:pionexch}. We do not consider orbital excitations, and assume the isospin symmetry $m_u=m_d$ throughout the paper.

\subsection{Center--of--Mass (CM) Correction}
\label{sec:cm}
In the independent-quark model, each quark moves in a confining potential centred at the origin. The three single-quark wave functions are then multiplied to construct the baryon wave function,
$\Phi=\psi_1\psi_2\psi_3$. This product wave function is a good approximation to how each quark moves relative to the centre of the well, but it has a drawback that is unrelated to the internal structure we are trying to describe. Each factor $\psi_i$ is obtained by solving a Dirac equation centred at the origin, independently of the other two quarks, and the resulting product wave function is not an eigenstate of the total momentum operator $\hat{\mathbf P}=\sum_i\hat{\mathbf p}_i$. A baryon at rest should have $\mathbf P=0$, but nothing in this construction forces the total momentum to be zero as the quarks move. Therefore, the sum of the three single-quark energies, $E_0=\sum_i E_i$, also includes the energy associated with the CM motion and needs to be corrected to obtain the baryon mass. The CM correction is particularly important for light quarks. Since they are more relativistic, their wave functions have a larger momentum spread, leading to a larger contribution from the CM motion. For heavy quarks, the momentum spread is smaller compared with their large mass, so the relative size of the correction is generally smaller.

To remove the CM contribution, we first need to estimate the momentum carried by each quark. For a Dirac radial wave function with upper component $g(r)$, corresponding to orbital angular momentum $l$, and lower component $f(r)$, with $l'=l+1$, the mean-square momentum of a confined quark can be written in terms of two radial integrals. The upper and lower components contribute separately, and there is no cross term between them:
\begin{align}
\langle p_q^2\rangle &= \int_0^\infty\!\left[\left(\frac{dg}{dr}\right)^{\!2}+\frac{l(l+1)}{r^2}g^2\right]dr
\nonumber\\
&\quad+\int_0^\infty\!\left[\left(\frac{df}{dr}\right)^{\!2}+\frac{l'(l'+1)}{r^2}f^2\right]dr
\label{eq:p2def}
\end{align}
Substituting the Gaussian solution of Eq.~\eqref{eq:gaussianansatz}, using the elementary Gaussian moment integrals $\int_0^\infty r^{2n}e^{-r^2/r_{0q}^2}\,dr=\tfrac12\Gamma(n+\tfrac12)\,r_{0q}^{2n+1}$, and eliminating $r_{0q},\eta_q$ in favor of $E_q',m_q'$ through $r_{0q}=(a\lambda_q)^{-1/4}$, $\eta_q=1/(r_{0q}\lambda_q)$ and the equation Eq.~\eqref{eq:cubic}, reduces the Eq.~\eqref{eq:p2def} to,
\begin{equation}
\langle p_q^2\rangle = \frac{(11E_q'+m_q')(E_q'^2-m_q'^2)}{6(3E_q'+m_q')}
\label{eq:p2}
\end{equation}
 This quantity grows with the confinement strength $a$. Larger $a$ squeezes the orbital to a smaller oscillator length $r_{0q}\propto a^{-1/4}$, and localizing a quark more tightly in position necessarily broadens its spread in momentum. 

To correct this unwanted CM contribution in the baryon mass calculation, we use the PY projection method. The method restores translational invariance exactly, at the level of the wave function itself, by explicitly projecting the product state $\Phi$ onto zero total momentum. 

The first step in implementing this correction is to determine how much the displaced state still overlaps with the original state if the entire state $\Phi$ is rigidly displaced by a vector $\mathbf R$ using the translation operator $\hat T(\mathbf R)=\exp(-i\mathbf R\cdot\hat{\mathbf P})$. This overlap for a single quark's orbital is
\begin{equation}
S_q(R) = \int \psi^\dagger(\mathbf r)\,\psi(\mathbf r-\mathbf R)\,d^3r
\label{eq:overlapdef}
\end{equation}
Inserting the spinor form of Eq.~\eqref{eq:spinoransatz}, the two upper-lower cross terms in $\psi^\dagger\psi$ vanish identically -- each is odd under $\hat r\to-\hat r$ inside the integrand, a direct consequence of the identity $\Omega_{1m}=-(\bm\sigma\cdot\hat r)\Omega_{-1m}$ relating the two spin-angle functions -- leaving only an upper-upper and a lower-lower piece. Substituting $\bm\rho=\mathbf r-\mathbf R/2$ gives $r^2+|\mathbf r-\mathbf R|^2=2\rho^2+R^2/2$, so the two displaced Gaussian of Eq.~\eqref{eq:gaussianansatz} collapse into a single Gaussian in $\bm\rho$ times an overall factor $e^{-R^2/4r_{0q}^2}$. The lower-lower piece additionally carries a factor $(\bm\sigma\cdot\hat r)(\bm\sigma\cdot\hat r')$ from the two spin-angle functions, which reduces to a scalar dot product once the term odd under the exchange is dropped by the same symmetry argument. Carrying this out for both pieces and adding them, this overlap can be worked out directly from the Gaussian solutions of Eq.~\eqref{eq:gaussianansatz},
\begin{equation}
S_q(R) = e^{-x_q}\,\frac{1+\eta_q^2\left(\tfrac32-x_q\right)}{1+\tfrac32\eta_q^2},
\qquad x_q=\frac{R^2}{4r_{0q}^2},\quad \eta_q=\frac{1}{r_{0q}\lambda_q}
\label{eq:overlap}
\end{equation}
which equals unity at $R=0$, as it should, and falls off smoothly once $R$ becomes comparable to the orbital's own size $r_{0q}$. A quark displaced by much more than its own confinement radius no longer resembles the original state at all. In the non-relativistic limit $\eta_q\to0$, Eq.~\eqref{eq:overlap} reduces to the plain Gaussian overlap $S_q(R)\to e^{-x_q}$ of a rigidly displaced wave packet, as it must once the relativistic lower component is switched off, and the $\eta_q^2$ term is entirely the correction generated by that lower component. Because the three quarks are confined independently, the analogous overlap for the whole baryon is the product of single-quark overlaps, $N(\mathbf R)=\prod_i S_{q_i}(R)$, which we refer to as the norm kernel. It measures how much of the original three-quark state survives a rigid displacement of the whole system by $\mathbf R$.

The second step is to turn this overlap information into a state of definite total momentum by the usual projection integral, $|\Psi(\mathbf P)\rangle = \int d^3R\,e^{i\mathbf P\cdot\mathbf R}\hat T(\mathbf R)|\Phi\rangle$, which superposes rigidly translated copies of $\Phi$ with plane-wave phases so that only the $\mathbf P=0$ piece of $\Phi$ survives when $\mathbf P=0$ is selected, while all values of $\mathbf P$ remain available for a general choice. The relative weight with which each total momentum $P=|\mathbf P|$ appears in $\Phi$ is exactly the three-dimensional Fourier transform of the norm kernel just constructed,
\begin{equation}
\widetilde N(\mathbf P) = \int d^3R\,e^{i\mathbf P\cdot\mathbf R}\,N(R)
\label{eq:FTdef}
\end{equation}
Since $N(\mathbf R)$ depends only on $R=|\mathbf R|$, the angular integral in Eq.~\eqref{eq:FTdef} can be done first, and the transform then reduces to a one-dimensional spherical Bessel integral, $\widetilde N(P)=4\pi\int_0^\infty dR\,R^2\,j_0(PR)\,N(R)$, and normalizing this to unit total probability gives the momentum-space density
\begin{equation}
\rho(P)=\frac{4\pi P^2\widetilde N(P)}{\int_0^\infty4\pi P^2\widetilde N(P)\,dP}
\label{eq:rhoP}
\end{equation}
Physically, $\rho(P)$ is the probability distribution of total momentum that the uncorrelated three-orbital state $\Phi$ already contains. It is this same spread that made $E_0=\sum_iE_i$ an overestimate of the rest mass in the first place.

Having the probability distribution for the baryon's total momentum in hand, we can now ask what rest mass is consistent with it, without needing to assume in advance how the CM kinetic energy splits off from the rest energy. A single relativistic particle of mass $M$ carrying momentum $P$ has energy $\sqrt{M^2+P^2}$. Averaging this over the distribution $\rho(P)$ and demanding that the result reproduce the total independent-quark energy $E_0$ fixes the projected mass $M_{\rm core}$ self-consistently,
\begin{equation}
\int_0^\infty \rho(P)\sqrt{M_{\rm core}^2+P^2}\,dP = E_0 = \sum_i E_i 
\label{eq:EPY}
\end{equation}
which is solved numerically to obtain the baryon core mass, $M_{\rm core}$.

\subsection{Residual interactions}
\label{sec:residual}
The confined three-quark core discussed earlier gives the gross mass scale of a baryon. However, the method cannot distinguish states with identical flavor content and different total spin, as the confining Hamiltonian of Eq.~\eqref{eq:H1} does not depend on spin. The fine and hyperfine splittings must therefore arise from residual QCD interactions that are not included in the mean confining field used to generate the single-quark orbitals. Two such contributions are included here. At short interquark distances, where the strong coupling is sufficiently moderate for a perturbative treatment, one gluon exchange between a pair of quarks produces both a spin-independent energy shift and a spin-dependent hyperfine contribution, discussed in Sec.~\ref{sec:gluon}. At longer distances, spontaneous chiral symmetry breaking supports the light $u,d$ quarks with a pion cloud, which contributes its own self-energy and exchange corrections. The resulting pion cloud contributes through self-energy and quark-exchange corrections, discussed in Sec.~\ref{sec:pion}. These pion contributions are absent, or strongly suppressed, for baryons composed entirely of heavy quarks. Both types of interactions are treated perturbatively on top of the PY-projected core mass $M_{\rm core}$ of Sec.~\ref{sec:cm}. Importantly, all residual contributions are evaluated using the same Dirac orbitals $g_q(r),f_q(r)$, that were obtained from the confining Hamiltonian and used in determining $M_{\rm core}$. The final physical baryon mass is then the sum of the projected core and the four residual contributions,
\begin{equation}
M_B = M_{\rm core} + \Delta E_g^{E} + \Delta E_g^{M} + \Delta E_\pi^{\rm self} +
\Delta E_\pi^{\rm exch}
\label{eq:massformula}
\end{equation}
Treating these four terms perturbatively is only justified if they are actually small compared to the core, and it is monitored explicitly through the residual-to-core ratio
\begin{equation}
\mathcal R = \frac{|\Delta E_\pi^{\rm self}|+|\Delta E_\pi^{\rm exch}|+|\Delta E_g^{E}|
+|\Delta E_g^{M}|}{|M_{\rm core}|}
\label{eq:pertratio}
\end{equation}
which is evaluated for every predicted state. For the baryons listed in the Tables~\ref{tab:calib}--\ref{tab:pred2}, $\mathcal R$ ranges from a few $\%$ for the heaviest, most non-relativistic triply heavy baryons -- where the perturbative treatment is on the firmest footing -- up to about $20\%$ for the lightest and most relativistic state considered, where the residual corrections are no longer small compared to the core. This does not by itself signal a failure of the model, as all the predicted masses agree with experiment and other theoretical models within $3\%$.

\subsubsection{One-gluon exchange}
\label{sec:gluon}
The first residual contribution arises from the exchange of a single gluon between two quarks, analogous to one-photon exchange between two electric charges in Quantum Electrodynamics (QED). The strength of this interaction is governed by the strong coupling constant, $\alpha_s$, the QCD counterpart of the fine-structure constant $\alpha\simeq1/137$ in QED. However, $\alpha_s$ is not a fixed number because gluons themselves carry color charge and interact with one another, and the effective strength of the interaction depends on the energy, or equivalently the momentum scale $\mu$, at which it is probed. This scale dependence is called the running of the coupling, and at short distances (large $\mu$) it makes $\alpha_s$ smaller, the property known as asymptotic freedom, which allows quarks and gluons to be treated perturbatively at these energies. At one-loop order, the scale dependence of the running coupling constant~\cite{Prosperi:2006hx} is governed by the equation
\begin{equation}
\frac{d\alpha_s}{d\ln\mu^2} = -\beta_0\frac{\alpha_s^2}{4\pi}
\label{eq:rge}
\end{equation}
where $\beta_0=(33-2N_f)/3$ is the one-loop beta-function coefficient and $N_f$ is the number of quark flavors light enough to be produced at the scale $\mu$. The negative sign on the right-hand side is the essential physics of Eq.~\eqref{eq:rge} because for  $\alpha_s^2>0$, the coupling can only decrease as $\mu$ increases, for any positive $\beta_0$. Substituting $t=\rm{ln}\mu^2$ in Equation~\eqref{eq:rge} and integrating $d\alpha_s/\alpha_s^2=-(\beta_0/4\pi)\,dt$ from a reference point $\mu_0$ up to $\mu$ gives
\begin{equation}
\frac{1}{\alpha_s(\mu^2)} = \frac{1}{\alpha_s(\mu_0^2)}+\frac{\beta_0}{4\pi}\ln\frac{\mu^2}{\mu_0^2}
\label{eq:rgeint}
\end{equation}
Defining $\Lambda_{\rm QCD}$ as the scale at which the right-hand side of Eq.~\eqref{eq:rgeint} vanishes eliminates the reference point $\mu_0$ in favor of $\Lambda_{\rm QCD}$ and gives the familiar running coupling
\begin{equation}
\alpha_s(\mu^2) = \frac{4\pi}{\beta_0\ln(\mu^2/\Lambda_{\rm QCD}^2)}
\label{eq:alpharun}
\end{equation}
Equation~\eqref{eq:alpharun} vanishes smoothly as $\mu\to\infty$ but diverges as $\mu\to\Lambda_{\rm QCD}$ (the Landau pole). That is exactly the regime relevant here, as the momentum exchanged between two quarks in a baryon is only a few hundred MeV, comparable to $\Lambda_{\rm QCD}$. So Eq.~\eqref{eq:alpharun} cannot be used directly, especially for light-quark pairs where perturbative gluon exchange inside a bound hadron is assumed. As the confining potential of Eq.~\eqref{eq:H1} already includes the dominant, non-perturbative part of the inter-quark force, the one-gluon exchange supplies only the residual, short-distance piece responsible for the fine and hyperfine splittings. We therefore use an infrared-finite version of Eq.~\eqref{eq:alpharun} that never diverges. For a pair of quarks $i,j$ with energies $E_i,E_j$, we evaluate the coupling at the reduced-energy scale $\mu_{ij}=E_iE_j/(E_i+E_j)$, with a fixed offset $\mu_{\rm IR}^2$ inside the logarithm,
\begin{equation}
\alpha_s(\mu_{ij}) =\!\frac{4\pi}{\beta_0\,
\ln\!\left[(\mu_{ij}^2+\mu_{\rm IR}^2)/\Lambda_{\rm QCD}^2\right]}
\label{eq:alphas}
\end{equation}
The offset keeps the logarithm's argument above $\mu_{\rm IR}^2/\Lambda_{\rm QCD}^2$, so the
Landau pole is avoided. $\mu_{\rm IR}$ sets the scale below which the coupling saturates rather than diverges. By using the three constant parameter values to be  
$\Lambda_{\rm QCD}=0.25~{\rm GeV},\quad \mu_{\rm IR}=0.70~{\rm GeV}$ and 
$\beta_0=9\ (N_f=3)$, the $\alpha_s(\mu_{ij})$ varies from 0.42 to 0.67 depending on the flavor of quark pairs. This range of $\alpha_s$ is similar to the value used by several papers from the original RIQ model.   

Although the running coupling fixes the overall strength of one-gluon exchange, the color structure of the interaction -- how strongly two quarks in a baryon attract or repel each other through gluon exchange also needs to be considered. Each quark carries a color charge described by the eight generators $T^a=\lambda^a/2$ ($a=1,\dots,8$) of the color SU(3) group, with $\lambda^a$ the Gell-Mann matrices and $\mathbf T_i$ denoting this set of color operators acting on quark $i$. Two quarks combine, as $\mathbf3\otimes\mathbf3= \bar{\mathbf3}\oplus\mathbf6$, where their color states can be combined either antisymmetrically forming a color-antitriplet $\bar{\mathbf 3}$, or symmetrically forming a color-sextet $\mathbf 6$. Only the antisymmetric $\bar{\mathbf 3}$ combination can subsequently combine with the third quark's color to form the baryon as a whole in a color singlet state. Using the identity $\mathbf T_i\!\cdot\!\mathbf T_j=\tfrac12[(\mathbf T_i+\mathbf T_j)^2-
\mathbf T_i^2-\mathbf T_j^2]$, together with $\mathbf T_i^2=C_F=4/3$ for the Casimir invariant of a single quark in the fundamental (triplet) representation and $(\mathbf T_i+\mathbf T_j)^2=C_{\bar{\mathbf3}}=4/3$ for the antitriplet pair, gives the pairwise color factor
\begin{equation}
\left\langle\frac{\lambda_i^a}2\frac{\lambda_j^a}2\right\rangle
= \mathbf T_i\!\cdot\!\mathbf T_j = \frac12\left(\frac43-\frac43-\frac43\right) = -\frac23
\label{eq:colorfactor}
\end{equation}
The negative sign means the color interaction is attractive, as required for gluon exchange to bind the baryon. The value $-2/3$ is the same for every pair in a color singlet state, independent of flavor. With the color factor fixed, the one-gluon-exchange potential between quarks $i,j$ at separation $r_{ij}=|\mathbf r_i-\mathbf r_j|$ splits into a spin-independent Coulomb piece and a spin-dependent contact piece,
\begin{align}
V_{ij}^{E}(r) &= -\frac{2}{3}\,\frac{\alpha_s(\mu_{ij})}{r_{ij}}, \label{eq:VE}\\
V_{ij}^{M} &= \frac{16\pi}{9}\,\frac{\alpha_s(\mu_{ij})}{E_iE_j}\,
\delta^{(3)}(\mathbf r_{ij})\,\bm\sigma_i\!\cdot\!\bm\sigma_j \label{eq:VM}
\end{align}
with the magnitude of the color factor of Eq.~\eqref{eq:colorfactor} entering both. In $V^E_{ij}$, the $1/r_{ij}$ falloff is the ordinary
Coulomb form, generated by exchange of the timelike component of the gluon field and independent of either quark's spin. In $V^M_{ij}$, the coefficient $16\pi/9$ comes from this same non-relativistic reduction applied to the spatial, current--current part of gluon exchange rather than its Coulomb part; as for the analogous magnetic (Breit) term in QED, this current--current reduction carries the opposite overall sign to the Coulomb term, so $V^M_{ij}$ comes out positive even though the same attractive color factor enters its magnitude. The factor $1/(E_iE_j)$ enters because the quarks are bound and moving, so their wave functions are normalized by their actual bound-state energies rather than their rest masses. The delta function $\delta^{(3)}(\mathbf r_{ij})$ confines the interaction to the point where the two quarks coincide, and $\bm\sigma_i\!\cdot\!\bm\sigma_j$ makes it depend on their relative spin alignment -- the only place in the gluon-exchange potential able to tell a spin-$\tfrac12$
baryon apart from its spin-$\tfrac32$ partner of identical flavor content. Taking
$\langle\Psi_B|V^{E,M}_{ij}|\Psi_B\rangle$ with the quark density
$\rho_i(r)=g_i^2(r)+f_i^2(r)$ and carrying out the angular integration gives the energy shifts
actually added to the baryon mass,

\begin{align}
\Delta E_g^{E} &= -\frac23\sum_{i<j}\alpha_s(\mu_{ij})\,I^E_{ij} \label{eq:dEE}\\
\Delta E_g^{M} &= \kappa_M\,\frac23\sum_{i<j}\alpha_s(\mu_{ij})
\left\langle\bm\sigma_i\!\cdot\!\bm\sigma_j\right\rangle I^M_{ij} \label{eq:dEM}
\end{align}
where the sum runs over the three quark pairs, and
\begin{equation}
I^E_{ij}\propto\int\!\!\int_0^\infty\! dr\,dr'\,r^2r'^2\frac{\rho_i(r)\rho_j(r')}{\max(r,r')}, 
\qquad
I^M_{ij}\propto\int d^3r\,\rho_i(r)\rho_j(r)
\end{equation}

$\kappa_M$ is an overall color--magnetic strength. It plays the role that a fixed coupling $\alpha_c=0.58$ played for both channels at once in the original light-baryon model~\cite{Barik:1986mq}. $I^E_{ij}$ and $I^M_{ij}$ are, respectively, the Coulomb-shell average and the coincident-point overlap already discussed for Eqs.~\eqref{eq:VE}--\eqref{eq:VM}.  Physically, $\Delta E_g^{E}$ shifts a whole flavor multiplet together, independent of spin, while $\Delta E_g^{M}$ is proportional to $\langle\bm\sigma_i\!\cdot\!\bm\sigma_j\rangle$ and
generates the hyperfine splitting between spin-$\tfrac12$ and spin-$\tfrac32$ partners of the same flavor content ($\Sigma_c$--$\Sigma_c^*$, $\Xi_c'$--$\Xi_c^*$, $\Omega_c$--$\Omega_c^*$,
and their $b$-flavored analogues), falling off with the pair's quark masses -- consistent with heavy-quark effective theory and known quarkonium hyperfine systematics~\cite{PhysRevD.47.3013}.

\subsubsection{Pion cloud and the quark--pion coupling}
\label{sec:pion}
The second residual effect is present only for the light $u,d$ quarks, with no counterpart for $s,c,b$. It follows from the approximate chiral symmetry of QCD. In the limit $m_u,m_d\to0$, the Lagrangian is invariant under independent rotations of the left- and right-handed quark fields, but the vacuum breaks this symmetry spontaneously. By the Goldstone theorem, this produces a massless boson, the pion. Because $m_u,m_d$ are not actually zero, the symmetry is only approximate, and the pion is instead unusually light, $m_\pi=0.1396$~GeV, well below the $\sim1$~GeV hadronic scale set by confinement.

This near-symmetry is encoded in the PCAC (partially conserved axial current) hypothesis. For the axial current $A^\mu_a=\bar q\gamma^\mu\gamma_5\tfrac{\tau_a}2 q$ carried by the light
quarks ($\tau_a$ the isospin Pauli matrices, $a=1,2,3$ for $\pi^0,\pi^+,\pi^-$), the approximate symmetry gives
\begin{equation}
\partial_\mu A^\mu_a=f_\pi m_\pi^2\phi_{\pi,a}
\end{equation}
with $\phi_{\pi,a}$ the pion field and $f_\pi=0.093$~GeV the pion decay constant. PCAC connects the axial coupling of a single confined quark to its coupling to pions, needed for the pion-cloud corrections below. 

The relevant quantity is the axial charge $g_A^{(q)}$ -- the single-quark version of the nucleon's axial coupling $g_A\simeq1.27$, the constant that measures how strongly a nucleon's spin couples to the weak axial current, and whose value is read directly off the neutron's beta-decay rate. It is computed directly from the same $1s_{1/2}$ orbital, with large and small components $g_q(r),f_q(r)$,
as the expectation value of the axial current's spatial part,
\begin{equation}
g_A^{(q)} = \int_0^\infty\!\left[g_q^2(r)-\tfrac13 f_q^2(r)\right]dr 
\label{eq:gA}
\end{equation}
The weight $-\tfrac13$ on $f_q(r)$, the same combination used in the MIT bag model's nucleon axial charge, means that any genuine relativistic contribution from $f_q(r)$ instead pulls $g_A^{(q)}$ below unity.

Evaluating Eq.~\eqref{eq:gA} for the Gaussian orbitals of Eq.~\eqref{eq:gaussianansatz}, with
the normalization $\int_0^\infty(g_q^2+f_q^2)\,dr=1$, gives
\begin{equation}
g_A^{(q)} = \frac{1-\tfrac12\eta_q^2}{1+\tfrac32\eta_q^2},
\label{eq:gAeta}
\end{equation}
and eliminating $\eta_q^2$ in favor of $E_q', m_q'$ via
$\eta_q^2 = (E_q'-m_q')\\/[3(E_q'+m_q')]$, reduces this to
\begin{equation}
g_A^{(q)} = \frac{1}{3}\cdot\frac{5E_q'+7m_q'}{3E_q'+m_q'} 
\label{eq:gAclosed}
\end{equation}
which returns $g_A^{(q)}\to1$ in the non--relativistic limit $E_q'\to m_q'$, and drops below unity as the quark becomes more relativistic. 

Apart from the axial charge $g_A^{(q)}$, which is a property of an isolated quark, we need the quark--pion coupling $f_{qq\pi}$. The two are related at the quark level as in bag-model
calculations~\cite{Duck:1976fp,Duck:1978mn,PhysRevD.21.1975},
\begin{equation}
f_{qq\pi} = g_{A,{\rm eff}}^{(q)}\,\frac{m_\pi}{2f_\pi} 
\label{eq:GT}
\end{equation}
where $g_{A,{\rm eff}}^{(q)}=(g_{A,{\rm quark\;scale}})\,g_A^{(q)}$, with
$g_{A,{\rm quark\;scale}}=1$ fixed at its natural value, so
$g_{A,{\rm eff}}^{(q)}=g_A^{(q)}$. This $f_{qq\pi}$ enters
both the self-energy and exchange terms below.

The interaction responsible for both effects is the quark--pion Lagrangian
\begin{equation}
\mathcal L_{q\pi}=\left(g_{A,{\rm eff}}^{(q)}/2f_\pi\right)
\bar\psi\gamma^\mu\gamma_5\tau^a\psi\,\partial_\mu\phi^a\,
\label{eq:Lqpi}
\end{equation}
where $\psi$ is now the full four-component confined-quark field and $\bar\psi\gamma^\mu\gamma_5\tau^a\psi$ is its axial current density, exactly the operator whose matrix element defines $g_A^{(q)}$ above. A quark can emit a virtual pion of momentum $\mathbf k$ and reabsorb it, lowering its own energy; summing this process over all possible pion momenta gives the pion self-energy correction
\begin{equation}
\Delta E_\pi^{\rm self} = -\frac{1}{3}\cdot\frac{1}{16\pi^3f_\pi^2}\int d^3k\,
\frac{|F_\pi(k)|^2 f_{qq\pi}^2}{\sqrt{k^2+m_\pi^2}}
\label{eq:pionself}
\end{equation}
summed over the $u,d$ quarks. The factor $\tfrac13$ is the isospin trace that results from averaging over the three possible pions
$\pi^0,\pi^+,\pi^-$ that can be emitted and reabsorbed, and $\sqrt{k^2+m_\pi^2}$ is simply the relativistic energy carried by a virtual pion of momentum $k=|\mathbf k|$. The remaining ingredient, $F_\pi(k)$, is the quark--pion transition form factor, now evaluated at general momentum $k$ and built, like every other quantity in this section, directly from the confined-quark orbitals,
\begin{equation}
F_\pi(k)=4\pi\int_0^\infty dr\,r^2j_0(kr)\left[g^2(r)-f^2(r)\right]\,
\label{eq:Fpi}
\end{equation}
the ordinary spherical Fourier transform of the axial-current density, with $j_0(kr)=\sin(kr)/kr$ the spherical Bessel function that appears in any such transform of a spherically symmetric source. Equation~\eqref{eq:pionself} is a single-quark self-energy where the emitting and reabsorbing quark stays inside the same baryon throughout, so the loop involves no intermediate baryon
state of different flavor content. 

The self-energy diagram involves only a single quark, but a pion emitted by one quark can equally well be absorbed by a different quark inside the same baryon, giving a genuine two-body pion-exchange force between them. Replacing the closed loop of Eq.~\eqref{eq:pionself} by the propagation of a real pion between quarks $i$ and $j$, with the static Yukawa propagator $D_\pi(r)=e^{-m_\pi r}/4\pi r$ whose exponential falloff sets the characteristic range of the force at $1/m_\pi\approx1.4$~fm, gives rise to the exchange energy,
\begin{equation}
\Delta E_\pi^{\rm exch} = -\frac{1}{18f_\pi^2}\sum_{i<j}
\left\langle\bm\sigma_i\!\cdot\!\bm\sigma_j\right\rangle
\left\langle\bm\tau_i\!\cdot\!\bm\tau_j\right\rangle \bar I^\pi_{ij}\, f_{qq\pi}^{(i)}f_{qq\pi}^{(j)}
\label{eq:pionexch}
\end{equation}
where the sum again runs over the three quark pairs in the baryon, $\bar I^\pi_{ij}=\tfrac12[I^\pi_{ii}+I^\pi_{jj}]$ is simply the average of the two quarks' own self-energy loop integrals of Eq.~\eqref{eq:pionself} (a convenient way of reusing quantities already computed, rather than a new radial integral), and $\bm\tau_i\!\cdot\!\bm\tau_j$ is the isospin analogue of the spin factor $\bm\sigma_i\!\cdot\!\bm\sigma_j$ already defined in Eq.~\eqref{eq:sigmadot}: it takes the value $\tau_f=-3$ for a pair of light quarks coupled to total isospin zero (the $u$--$d$ pair inside $\Lambda_c$, for instance) and $\tau_f=+1$ for a pair coupled to total isospin one (as in $\Sigma_c$). Combining the two factors, the product $\langle\bm\sigma_i\!\cdot\!\bm\sigma_j\rangle\tau_f$ works out to $(-3)(-3)=+9$ for the spin-singlet,
isospin-singlet $\Lambda_c$-type configuration, but only $(+1)(+1)=+1$ for the spin-triplet,
isospin-triplet $\Sigma_c$-type configuration -- so pion exchange alone binds the $\Lambda_c$-type
pair roughly an order of magnitude more strongly than the $\Sigma_c$-type pair, before any
difference in the loop integrals or couplings themselves is even taken into account. This is exactly the same spin-isospin channel
that dominates the long-range part of the nuclear force between nucleons, appearing here at the quark level instead. Because the isospin matrices $\tau^a$ act only on $u,d$ quarks, both
pion terms of Eqs.~\eqref{eq:pionself} and \eqref{eq:pionexch} vanish identically whenever a given light pair is replaced by a strange quark, or whenever the baryon contains
no light quark at all, and both terms shrink steadily as light quarks are replaced by $c$ or $b$ quarks across the baryon spectrum.

\section{Results and discussion}
\label{sec:results}
\begin{figure*}[tb]
\centering
\includegraphics[width=0.8\textwidth]{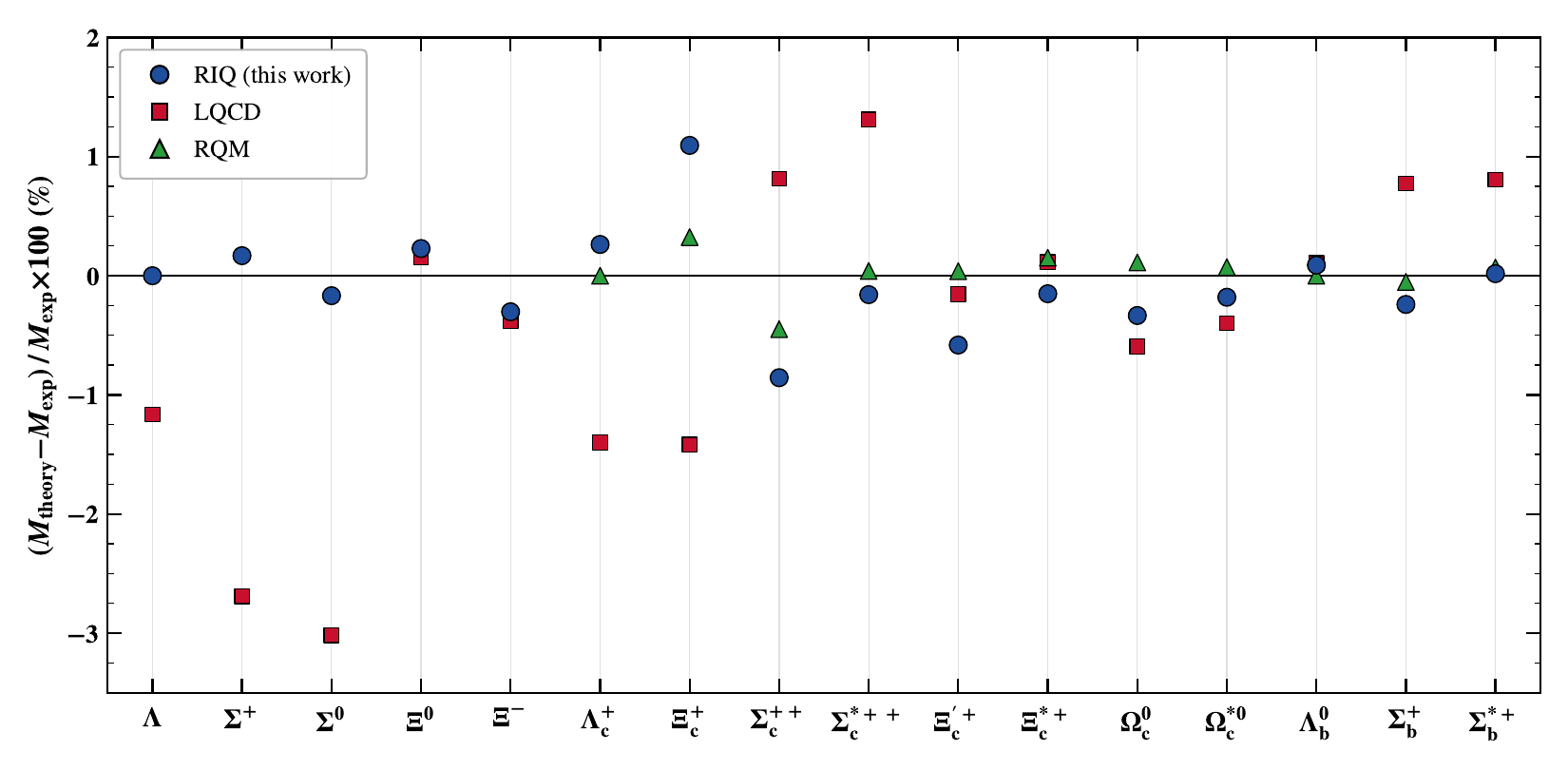}
\caption{Percentage deviation $(M_{\rm theory}-M_{\rm exp})/M_{\rm exp}\times100$ of the RIQ model (this work), lattice-QCD, and relativistic quark-diquark-model masses from the experimentally measured values, for the sixteen baryon states of Table~\ref{tab:calib}. RIQ predicted masses are within about $\pm 1.1\%$ of the experimental values.}
\label{fig:deviation}
\end{figure*}

\begin{figure*}[tb]
\centering
\includegraphics[width=0.8\textwidth]{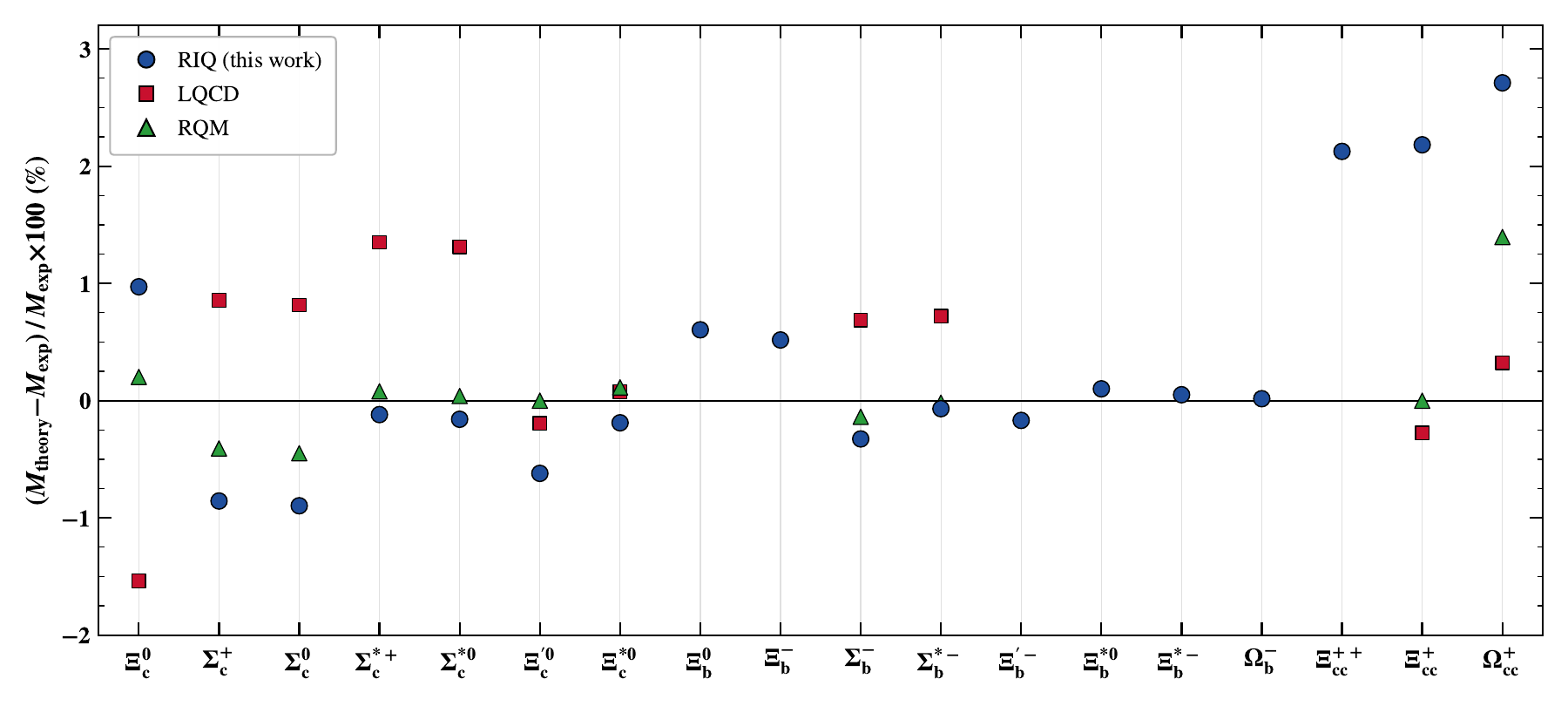}
\caption{Percentage deviation $(M_{\rm theory}-M_{\rm exp})/M_{\rm exp}\times100$ of the RIQ
(this work), lattice-QCD, and relativistic quark-diquark model
masses from the experimentally measured values, for the eighteen states of Table~\ref{tab:pred1} that have a measured mass but were not used for optimizing input parameters. RIQ deviations remain within about $\pm 1.0\%$ for the fifteen singly-heavy states, and rise to $+2.1$--$2.7\%$ for other states.}
\label{fig:deviation-pred}
\end{figure*}

The confinement and constituent-mass input parameters used in the model $(a, V_0, m_u{=}m_d, m_s, m_c, m_b)$ are optimized to predict the sixteen calibration baryon masses as close as possible to the experimentally measured values. Table~\ref{tab:params} shows the best-fit parameter values. Table~\ref{tab:calib} compares our calibrated sixteen baryon masses ($M_{\rm RIQ}$) with the experimental values ($M_{\rm exp}$)~\cite{PDG2026Listings}. The signed differences $\Delta M=M_{\rm RIQ}-M_{\rm exp}$ remain below about $1.1\%$ for every state, the largest deviation ($+1.087\%$) occurring for $\Xi_c^+$. We compare our results with the mass predictions from lattice QCD (LQCD) ($M_{\rm LQCD}$) ~\cite{Durr:2008zz,Brown:2014ena} and the mass from the relativistic quark-diquark model (RQM) ($M_{\rm RQM}$)~\cite{Ebert:2002ig,Faustov:2021qqf,Ebert:2011kk}. Table~\ref{tab:calib} shows that our results are in good agreement with different theoretical predictions and experimental data.  Tables~\ref{tab:pred1} and~\ref{tab:pred2} list the remaining ground state baryons, none of which are used in the optimization process of input parameters. The eighteen states of Table~\ref{tab:pred1} have an experimentally measured mass~\cite{PDG2026Listings} and serve as a genuine test of our optimized model parameters. We compare our results with LQCD~\cite{Brown:2014ena} and the RQM model predictions~\cite{Ebert:2011kk,Ebert:2002ig}. Table~\ref{tab:pred2}  extends the study to additional ground-state baryons whose masses have not yet been measured experimentally, including unobserved states from the same flavor sectors as well as mixed-$bc$ and triply heavy baryons and compares them with LQCD~\cite{Brown:2014ena} and the RQM model~\cite{Ebert:2002ig,Ebert:2011kk,Faustov:2021qqf} available in the literature. Figure~\ref{fig:deviation} compares the predictions of the RIQ model, LQCD, and the RQM model with the experimental masses of the sixteen calibration baryons. The percentage deviation, defined as $(M_{\rm theory}-M_{\rm exp})/M_{\rm exp}\times100$, where $M_{\rm theory}$ is the predicted mass, is shown for each baryon. The RIQ predictions (filled circles) track the experimental masses \cite{PDG2026Listings} closely across the full light-to-bottom mass range, with deviations comparable in size to, and for several states smaller than, those of the lattice-QCD and quark-diquark-model calculations. Figure~\ref{fig:deviation-pred} shows the same comparison for the eighteen measured states of Table~\ref{tab:pred1}, confirming our model agreement holds beyond the calibrated baryons themselves.

The projected core mass, $M_{\rm core}$, provides the dominant contribution to the total baryon mass, while the residual interactions generate spectroscopic splittings. The relative importance of these corrections varies across the different flavor sectors. The pion contributions become increasingly suppressed as the light-quark content decreases and vanish for baryons composed entirely of heavy quarks. Therefore, the relative contributions of the gluon and pion terms provide additional insight into the mass systematics across the light, singly heavy, doubly heavy, and triply heavy sectors.

\begin{table}[t]
\centering
\caption{The six optimized input parameters used for the prediction of baryon masses. All dimensionful quantities are in GeV, with $a$ in GeV$^3$.}
\label{tab:params}
\footnotesize
\begin{tabular}{@{}llc@{}}
\toprule
Parameter & Meaning & Value \\
\midrule
$a$          & confinement strength & $0.01307$ \\
$V_0$        & potential offset     & $-0.04707$ \\
$m_u=m_d$    & light quark mass     & $0.23986$ \\
$m_s$        & strange quark mass   & $0.38029$ \\
$m_c$        & charm quark mass     & $1.68428$ \\
$m_b$        & bottom quark mass    & $5.05546$ \\
\bottomrule
\end{tabular}
\end{table}

\begin{table*}[tb]
\centering
\caption{The table shows a comparison between different theoretical model predictions and the experimentally measured ground-state baryon masses (all in GeV unless noted). $\Delta M = M_{\rm RIQ}-M_{\rm exp}$ is the difference between the RIQ model prediction and the measured ground-state baryon masses. }
\label{tab:calib}
\footnotesize
\begin{tabular}{@{}lcccccccc@{}}
\toprule
Baryon & Quarks & $J^P$ & $M_{\rm exp}$ & $M_{\rm RIQ}$ & $\Delta M$ (MeV) & Diff.\ (\%) & $M_{\rm LQCD}$ & $M_{\rm RQM}$ \\
\midrule
$\Lambda$        & $uds$ & $1/2^+$ & 1.116 & 1.116 & $0$  & $+0.000$ & $1.103$ & --   \\
$\Sigma^+$       & $uus$ & $1/2^+$ & 1.189 & 1.191 & $+2$  & $+0.168$ & $1.157$ & --   \\
$\Sigma^0$       & $uds$ & $1/2^+$ & 1.193 & 1.191 & $-2$  & $-0.168$ & $1.157$ & --   \\
$\Xi^0$          & $ssu$ & $1/2^+$ & 1.315 & 1.318 & $+3$  & $+0.228$ & $1.317$ & --   \\
$\Xi^-$          & $ssd$ & $1/2^+$ & 1.322 & 1.318 & $-4$  & $-0.303$ & $1.317$ & --   \\
$\Lambda_c^+$    & $udc$ & $1/2^+$ & 2.286 & 2.292 & $+6$  & $+0.262$ & $2.254$ & $2.286$  \\
$\Xi_c^+$        & $usc$ & $1/2^+$ & 2.468 & 2.495 & $+27$ & $+1.094$ & $2.433$ & $2.476$  \\
$\Sigma_c^{++}$  & $uuc$ & $1/2^+$ & 2.454 & 2.433 & $-21$ & $-0.856$ & $2.474$ & $2.443$  \\
$\Sigma_c^{*++}$ & $uuc$ & $3/2^+$ & 2.518 & 2.514 & $-4$  & $-0.159$ & $2.551$ & $2.519$  \\
$\Xi_c'^{+}$     & $usc$ & $1/2^+$ & 2.578 & 2.563 & $-15$ & $-0.582$ & $2.574$ & $2.579$  \\
$\Xi_c^{*+}$     & $usc$ & $3/2^+$ & 2.645 & 2.641 & $-4$  & $-0.151$ & $2.648$ & $2.649$  \\
$\Omega_c^0$     & $ssc$ & $1/2^+$ & 2.695 & 2.686 & $-9$  & $-0.334$ & $2.679$ & $2.698$  \\
$\Omega_c^{*0}$  & $ssc$ & $3/2^+$ & 2.766 & 2.761 & $-5$  & $-0.181$ & $2.755$ & $2.768$  \\
$\Lambda_b^0$    & $udb$ & $1/2^+$ & 5.620 & 5.625 & $+5$  & $+0.089$ & $5.626$ & $5.620$  \\
$\Sigma_b^+$     & $uub$ & $1/2^+$ & 5.811 & 5.797 & $-14$ & $-0.241$ & $5.856$ & $5.808$  \\
$\Sigma_b^{*+}$  & $uub$ & $3/2^+$ & 5.830 & 5.831 & $+1$  & $+0.017$ & $5.877$ & $5.834$  \\
\bottomrule
\end{tabular}
\end{table*}

\begin{table*}[tb]
\centering
\caption{States that were not used for optimizing the input parameters, but have an experimentally measured mass (all masses in GeV unless noted).}
\label{tab:pred1}
\footnotesize
\begin{tabular}{@{}lcccccccc@{}}
\toprule
Baryon & Quarks & $J^P$ & $M_{\rm exp}$ & $M_{\rm RIQ}$ & $\Delta M$ (MeV) & Diff.\ (\%) & $M_{\rm LQCD}$ & $M_{\rm RQM}$ \\
\midrule
$\Xi_c^0$          & $dsc$ & $1/2^+$ & $2.471$ & 2.495 & $+24$ & $+0.971$ & $2.433$ & $2.476$  \\
$\Sigma_c^+$       & $udc$ & $1/2^+$ & $2.453$ & 2.432 & $-21$ & $-0.856$ & $2.474$ & $2.443$  \\
$\Sigma_c^0$       & $ddc$ & $1/2^+$ & $2.454$ & 2.432 & $-22$ & $-0.896$ & $2.474$ & $2.443$  \\
$\Sigma_c^{*+}$    & $udc$ & $3/2^+$ & $2.517$ & 2.514 & $-3$  & $-0.119$ & $2.551$ & $2.519$  \\
$\Sigma_c^{*0}$    & $ddc$ & $3/2^+$ & $2.518$ & 2.514 & $-4$  & $-0.159$ & $2.551$ & $2.519$  \\
$\Xi_c'^{0}$       & $dsc$ & $1/2^+$ & $2.579$ & 2.563 & $-16$ & $-0.620$ & $2.574$ & $2.579$  \\
$\Xi_c^{*0}$       & $dsc$ & $3/2^+$ & $2.646$ & 2.641 & $-5$  & $-0.189$ & $2.648$ & $2.649$  \\
$\Xi_b^0$          & $usb$ & $1/2^+$ & $5.792$ & 5.827 & $+35$ & $+0.604$ & -- & --   \\
$\Xi_b^-$          & $dsb$ & $1/2^+$ & $5.797$ & 5.827 & $+30$ & $+0.518$ & -- & --   \\
$\Sigma_b^-$       & $ddb$ & $1/2^+$ & $5.816$ & 5.797 & $-19$ & $-0.327$ & $5.856$ & $5.808$  \\
$\Sigma_b^{*-}$    & $ddb$ & $3/2^+$ & $5.835$ & 5.831 & $-4$  & $-0.069$ & $5.877$ & $5.834$  \\
$\Xi_b'^{-}$       & $dsb$ & $1/2^+$ & $5.935$ & 5.925 & $-10$  & $-0.168$ & -- & --   \\
$\Xi_b^{*0}$       & $usb$ & $3/2^+$ & $5.952$ & 5.958 & $+6$  & $+0.101$ & -- & --   \\
$\Xi_b^{*-}$       & $dsb$ & $3/2^+$ & $5.955$ & 5.958 & $+3$  & $+0.050$ & -- & --   \\
$\Omega_b^-$       & $ssb$ & $1/2^+$ & $6.046$ & 6.047 & $+1$  & $+0.017$ & -- & --   \\
$\Xi_{cc}^{++}$    & $ccu$ & $1/2^+$ & $3.622$ & 3.699 & $+77$ & $+2.126$ & -- & --   \\
$\Xi_{cc}^{+}$     & $ccd$ & $1/2^+$ & $3.620$ & 3.699 & $+79$ & $+2.182$ & $3.610$ & $3.620$  \\
$\Omega_{cc}^{+}$  & $ccs$ & $1/2^+$ & $3.726$ & 3.827 & $+101$ & $+2.711$ & $3.738$ & $3.778$  \\
\bottomrule
\end{tabular}
\end{table*}

\begin{table*}[t]
\centering
\caption{States predicted by theoretical models but not yet observed
experimentally. All masses are in GeV.
The symbol ``--'' indicates that no prediction is available.}
\label{tab:pred2}
\footnotesize
\setlength{\tabcolsep}{3pt}
\renewcommand{\arraystretch}{1.15}
\begin{tabular}{@{}lccccc!{\hspace{15pt}\vrule width 0.8pt\hspace{15pt}}lccccc@{}}
\toprule
Baryon & Quarks & $J^P$ & $M_{\rm RIQ}$ &
$M_{\rm LQCD}$ & $M_{\rm RQM}$ &
Baryon & Quarks & $J^P$ & $M_{\rm RIQ}$ &
$M_{\rm LQCD}$ & $M_{\rm RQM}$ \\
\midrule
$\Sigma_b^0$ & $udb$ & $1/2^+$ & 5.797 & 5.856 & 5.808 &
$\Xi_{bc}^{0}$ & $bcd$ & $1/2^+$ & 7.073 & 6.943 & 6.933 \\
$\Sigma_b^{*0}$ & $udb$ & $3/2^+$ & 5.831 & 5.877 & 5.834 &
$\Xi_{bc}'^{+}$ & $bcu$ & $1/2^+$ & 7.033 & 6.959 & 6.963 \\
$\Xi_b'^{0}$ & $usb$ & $1/2^+$ & 5.925 & -- & -- &
$\Xi_{bc}'^{0}$ & $bcd$ & $1/2^+$ & 7.033 & 6.959 & 6.963 \\
$\Omega_b^{-}$ & $ssb$ & $3/2^+$ & 6.078 & 6.085 & 6.088 &
$\Xi_{bc}^{+}$ & $bcu$ & $3/2^+$ & 7.096 & 6.985 & 6.980 \\
$\Xi_{cc}^{*++}$ & $ccu$ & $3/2^+$ & 3.781 & 3.692 & 3.727 &
$\Xi_{bc}^{*0}$ & $bcd$ & $3/2^+$ & 7.096 & 6.985 & 6.980 \\
$\Xi_{cc}^{*+}$ & $ccd$ & $3/2^+$ & 3.781 & 3.692 & 3.727 &
$\Omega_{bc}^{0}$ & $bcs$ & $1/2^+$ & 7.196 & 6.998 & 7.088 \\
$\Omega_{cc}^{*+}$ & $ccs$ & $3/2^+$ & 3.902 & 3.822 & 3.872 &
$\Omega_{bc}'^{0}$ & $bcs$ & $1/2^+$ & 7.160 & 7.032 & 7.116 \\
$\Xi_{bb}^{0}$ & $bbu$ & $1/2^+$ & 10.368 & 10.143 & 10.202 &
$\Omega_{bc}^{*0}$ & $bcs$ & $3/2^+$ & 7.218 & 7.059 & 7.130 \\
$\Xi_{bb}^{-}$ & $bbd$ & $1/2^+$ & 10.368 & 10.143 & 10.202 &
$\Omega_{ccc}^{++}$ & $ccc$ & $3/2^+$ & 5.053 & 4.796 & 4.712 \\
$\Xi_{bb}^{*0}$ & $bbu$ & $3/2^+$ & 10.402 & 10.178 & 10.237 &
$\Omega_{bbb}^{-}$ & $bbb$ & $3/2^+$ & 15.013 & 14.366 & 14.468 \\
$\Xi_{bb}^{*-}$ & $bbd$ & $3/2^+$ & 10.402 & 10.178 & 10.237 &
$\Omega_{ccb}^{+}$ & $ccb$ & $1/2^+$ & 8.360 & 8.007 & 7.984 \\
$\Omega_{bb}^{-}$ & $bbs$ & $1/2^+$ & 10.494 & 10.273 & 10.359 &
$\Omega_{ccb}^{*+}$ & $ccb$ & $3/2^+$ & 8.374 & 8.037 & 7.999 \\
$\Omega_{bb}^{*-}$ & $bbs$ & $3/2^+$ & 10.525 & 10.308 & 10.389 &
$\Omega_{cbb}^{0}$ & $bbc$ & $1/2^+$ & 11.677 & 11.195 & 11.198 \\
$\Xi_{bc}^{+}$ & $bcu$ & $1/2^+$ & 7.073 & 6.943 & 6.933 &
$\Omega_{cbb}^{*0}$ & $bbc$ & $3/2^+$ & 11.691 & 11.229 & 11.217 \\
\bottomrule
\end{tabular}
\end{table*}

\section{Summary and Conclusions}
\label{sec:summary}

We presented a heavy--flavor extension of the Barik--Dash relativistic independent-quark model, in which three independently confined, positive-energy Dirac
quarks are combined to form a baryonic state. The confining Dirac dynamics and the perturbative residual-interaction philosophy of the original RIQ model are retained at the single-quark level. The explicit momentum projection, the infrared-regulated running coupling, one-gluon exchange and the pion-cloud sector are specific to the present calculation.
The results are accordingly best regarded as predictions of an extended, RIQ-inspired model rather than a direct reproduction of the original construction. The spurious translational
motion is removed by a self-consistent numerical Peierls--Yoccoz projection of the full
three-orbital overlap kernel. The corrections from pion self-energy, pion-exchange, and pairwise color-electric/magnetic one-gluon-exchange corrections are added perturbatively to
the projected core mass. A six-parameter optimization is performed to predict the masses for the full doubly heavy, mixed-$bc$, and triply heavy ground-state spectra. The resulting spectrum has been compared with LQCD and RQM predictions, and experimental measurements where available. Our results are consistent with the predictions from these two theoretical models as well as experimentally measured values. Where the experimental baryon masses are not available, our predictions could be tested in future experiments.

\begin{acknowledgements}
We acknowledge the support of the National Institute of Science Education and Research (NISER), Bhubaneswar, and the Homi Bhabha National Institute (HBNI), Mumbai. We also acknowledge the support of the Department of Atomic Energy (DAE), Government of India. We thank Professors N. Barik, P. C. Dash, and S. Kar for their valuable discussions on the RIQ framework.
\end{acknowledgements}

\bibliographystyle{spphys}
\bibliography{bibliography}

\end{document}